\documentclass[prb,letterpaper,aps,superscriptaddress,floatfix,twocolumn]{revtex4-1}
\usepackage{graphicx}
\usepackage{amsmath}
\usepackage{subfigure}
\newcommand{\beq}{\begin{equation}}
\newcommand{\eeq}{\end{equation}}
\newcommand{\bk}{{{\bf{k}}}}

\newcommand{\bQ}{{{\bf{Q}}}}
\newcommand{\br}{{{\bf{r}}}}

\newcommand{\bq}{{\bf{q}}}

\newcommand{\beqa}{\begin{eqnarray}}
\newcommand{\eeqa}{\end{eqnarray}}
\newcommand{\pdg}{{\vphantom \dag}}
\newcommand{\dg}{{\dag}}
 
\newcommand{\bsigma}{{\boldsymbol \sigma}}
\newcommand{\bkappa}{{\boldsymbol \kappa}}
\newcommand{\btau}{{\boldsymbol \tau}}

\newcommand{\upa}{\uparrow}
\newcommand{\da}{\downarrow}

\begin{document}
\title{Superconductivity in Weyl metals}
\author{G. Bednik}
\affiliation{Department of Physics and Astronomy, University of Waterloo, Waterloo, Ontario 
N2L 3G1, Canada}
\author{A.A. Zyuzin}
\affiliation{Department of Physics, University of Basel, Klingelbergstrasse 82, CH-2056 Basel, Switzerland}
\author{A.A. Burkov}
\affiliation{Department of Physics and Astronomy, University of Waterloo, Waterloo, Ontario 
N2L 3G1, Canada} 
\affiliation{National Research University ITMO, Saint Petersburg 197101, Russia}
\date{\today}
\begin{abstract}
We report on a study of intrinsic superconductivity in a Weyl metal, i.e. a doped Weyl semimetal. 
Two distinct superconducting states are possible in this system in principle: a zero-momentum pairing 
BCS state, with point nodes in the gap function; and a finite-momentum FFLO-like state, with a full nodeless gap.
We find that, in an inversion-symmetric Weyl metal the odd-parity BCS state has a lower energy than the FFLO state, 
despite the nodes in the gap. The FFLO state, on the other hand, may have a lower energy in a noncentrosymmetric Weyl metal, 
in which Weyl nodes of opposite chirality have different energy. However, realizing the FFLO state is in general very 
difficult since the paired states are not related by any exact symmetry, which precludes a weak-coupling superconducting instability. 
We also discuss some of the physical properties of the nodal BCS state, in particular Majorana and Fermi arc surface states. 
\end{abstract}
\maketitle
\section{Introduction}
\label{sec:1}
The study of the interplay of superconductivity and nontrivial electronic structure topology has a long history, dating back to the 
work on superfluid $^3$He.~\cite{Volovik88,Volovik03,Volovik07} 
The recent discovery of topological insulators (TI)~\cite{Hasan10,Qi11} has reinvigorated this subject. 
Proximity-induced superconductivity on a 3D TI surface has been proposed as a route to realize Majorana fermions,~\cite{Fu08}
and bulk topological superconductivity has also been studied extensively.~\cite{Ludwig08,Hughes09,Fu10}

Most recently, Weyl, and closely related Dirac semimetals, first discovered theoretically~\cite{Wan11,Ran11,Burkov11-1,Xu11,Kane12,Fang12,Fang13,Bernevig15-1,Hasan15-5} 
and lately realized experimentally,~\cite{Hasan15-3,Fang15-2,Hasan15-4,Lu15,Hasan15-2,Fang15-1,Hasan15-1,Borisenko14,Shen14,Neupane14,Hasan15-6,Ando11,Kharzeev14,Gu15}
have come to the forefront of research on topologically-nontrivial phases of matter. 
Perhaps the most interesting new feature that Weyl semimetals bring to the subject is that they are gapless. The realization that gapless states of matter may be topologically 
nontrivial, just as gapped insulators, is of significant importance. 

The nontrivial electronic structure topology of Weyl semimetals arises from points of contact of nondegenerate conduction and valence bands, which act as monopole 
sources of Berry curvature and thus carry an integer topological charge.
A semimetal is realized when the Fermi level coincides with the points of contact and the valence band is filled while the conduction band is empty. 
This situation, however, is nongeneric and is unstable to impurities which will always give rise to unintentional doping, shifting the Fermi level into the conduction 
or valence bands. 
This motivates the study of {\em Weyl metals}, i.e. lightly-doped Weyl semimetals. It turns out that Weyl metals share most of the topologically-nontrivial 
properties with undoped Weyl semimetals.~\cite{Burkov14-2,Burkov14-3,Burkov15-1}
Two features of the Weyl metals, which are directly related to the Weyl node topology, are very important in this regard. 
One is that, when the Fermi energy is close enough to the Weyl nodes, the Fermi surface breaks up into disconnected sheets, each enclosing one Weyl node. 
The flux of the Berry curvature through each such Fermi surface sheet is equal to the topological charge of the corresponding node, which endows each 2D Fermi surface 
sheet with a nontrivial topological invariant, a Chern number. 
The second important property of a Weyl metal is the linearity of the band dispersion, which necessarily arises in the momentum-space vicinity of each topological charge, as 
follows from the so-called Atiyah-Bott-Shapiro construction.~\cite{Horava05} 
This property may be viewed as an emergent low-energy {\em chiral symmetry}, which is characteristic of Weyl metals. 

A nearly universal property of metals is the superconducting instability, which always exists, at least in nonmagnetic metals, at low enough temperature. 
From this viewpoint, the question of superconductivity in Weyl metals comes up naturally. 
Moreover, it is certain in this case that the superconductivity will be unconventional since, at least at weak coupling, the pairing must occur between states in a single 
nondegenerate conduction or valence band. 

As has been shown in Refs.~\onlinecite{Moore12,Aji14}, two distinct superconducting states are possible in Weyl metals in principle: a BCS (Bardeen-Cooper-Schrieffer) state, 
in which pairing occurs between momentum eigenstates, related to each other by inversion symmetry [assuming time reversal (TR) symmetry is broken, but inversion exists]; and an FFLO (Fulde-Ferrell-Larkin-Ovchinnikov)-like state, with finite-momentum pairs, where the states on the opposite sides of each Fermi surface sheet, enclosing individual Weyl nodes, are paired. 
In the latter case, the states which are paired are not related to each other by any exact symmetry. What makes such a superconducting state possible in principle, even at 
weak (but not infinitesimal) coupling, is the low-energy chiral symmetry of Weyl metals, mentioned above. 
Moreover, if the inversion symmetry is violated, and as a result, Weyl nodes of opposite chirality are shifted away from each other in energy to a significant degree,~\cite{Zyuzin12-2}
the FFLO state becomes the only superconducting state possible, since the existence of the BCS state relies on inversion symmetry. 

In this paper we show that, in the presence of inversion symmetry, the BCS state has a lower energy, at least for short-range momentum-independent pairing interaction and 
in the class of Weyl metals, considered in this paper. 
This conclusion differs from that, made in previous work on this subject,~\cite{Moore12,Aji14} which argued that the FFLO state has a lower energy. 
We will explain the origin of this disagreement below. 
For other recent work on Weyl superconductors, see Refs.~\onlinecite{Meng12,Thomale14,Tanaka15}. 

The rest of the paper is organized as follows. 
In Section~\ref{sec:2} we introduce the model of a Weyl metal we will use, which is based on the TI-NI (normal insulator) multilayer heterostructure, 
introduced by one of us before.~\cite{Burkov11-1}
This model has the benefit of being very simple and thus amenable to analytic calculations, yet more realistic than a generic low-energy model of a Weyl 
metal would be. In particular, it does not suffer from spurious exact chiral symmetry of the simplest ``relativistic" low energy models.  
We also derive in this section the BCS pairing Hamiltonian, the form of which plays an important role in our analysis. 
In Section~\ref{sec:3} we analyze the BCS Hamiltonian, derived in Section~\ref{sec:2}, using the standard mean-field theory and demonstrate 
explicitly that, in the presence of inversion symmetry the BCS pairing state has a lower energy than the FFLO state. 
In Section~\ref{sec:4} we discuss some of the physical properties of the nodal BCS state in more detail, in particular the topological properties of the mean-field Hamiltonian 
and the edge states. 
We discuss the differences between our results and previous work in Section~\ref{sec:5} and conclude by pointing out that, when the inversion symmetry (in addition to TR) is violated, the BCS state is destroyed while the FFLO state may survive. 

\section{Model and derivation of the BCS and FFLO Hamiltonians}
\label{sec:2}
As mentioned above, we will start from the model of a Weyl metal, based on the TI-NI multilayer heterostructure, introduced by one of us before.~\cite{Burkov11-1} 
The model has been described extensively in the literature and here we will only mention the most essential points, which are necessary to understand what follows. 
The momentum space Hamiltonian, describing the multilayer structure, is given by
\beq
\label{eq:1}
H_0 = v _F \tau^z (\hat z \times \bsigma) \cdot \bk + \hat t(k_z). 
\eeq
Here $\hat z$ is the growth direction of the heterostructure, $\bsigma$ are Pauli matrices, describing the real spin degree of freedom, while $\btau$ is the pseudospin, describing the 
top (T) and bottom (B) surfaces of the TI layers in the heterostructure, and $\hbar =1$ units are used here and throughout the paper.
 The operator $\hat t(k_z)$ describes the motion of the electrons in the growth direction of the structure. Explicitly it is 
given by
\beq
\label{eq:2}
\hat t(k_z) = t_S \tau^x + \frac{t_D}{2} \left( \tau^+ e^{i k_z d} + h.c. \right), 
\eeq
where $t_{S,D}$ are amplitudes for tunnelling between top and bottom surfaces of the same (S) or neighbouring (D) TI layers and $d$ is the superlattice period. 
We will take both $t_{S,D}$ to be positive for concreteness, this choice does not affect any of the physics. 
This structure describes a Dirac semimetal when $t_S = t_D$, with two Weyl fermion components of opposite chirality residing at the same point $(0, 0, \pi/d)$ in the first Brillouin zone (BZ). 
To create a Weyl semimetal, we need to separate the Weyl fermions of opposite chirality in momentum space. This may be accomplished by breaking either TR or 
inversion~\cite{Halasz12} symmetries.
We choose to break TR and (for now) keep inversion symmetry intact, as this creates the simplest kind of Weyl semimetal state with only two nodes. 
Breaking of TR is accomplished by adding a term $b \, \sigma^z$ to the Hamiltonian Eq.~\eqref{eq:1}
\beq
\label{eq:3}
H_0 = v _F \tau^z (\hat z \times \bsigma) \cdot \bk + \hat t(k_z) + b \sigma^z. 
\eeq
Physically this may arise, for example, from magnetized transition-metal impurities, introduced into the sample. 
Throughout this paper we will assume that undoped heterostructure is almost a Dirac semimetal, i.e. $|t_S - t_D|$ is small. 

We now introduce the simplest and most natural (at least for phonon-mediated pairing) kind of pairing interaction, i.e. an $s$-wave short-range, and thus necessarily singlet, pairing term, which 
in second-quantized notation has the form
\beq
\label{eq:4}
H_{int} = -U \int \,d^3 r \, \Psi^\dg_\upa(\br) \Psi^\dg_\da(\br) \Psi^\pdg_\da(\br) \Psi^\pdg_\upa(\br), 
\eeq
where $U > 0$. 
We want to eventually rewrite $H_{int}$ in the basis of the eigenstates of $H_0$. 
To this end, we first write the electron field operators in the following way
\beq
\label{eq:5}
\Psi^\dg_{\sigma}(\br) = \frac{1}{\sqrt{L_x L_y}} \sum_{\bk i \tau} \phi^*_{i \tau}(z) e^{- i \bk \cdot \br} c^\dg_{\bk i \sigma \tau}. 
\eeq
Here $i$ labels the unit cells of the TI-NI superlattice in the growth direction, $\sigma$ and $\tau$ are the spin and pseudospin labels, $\phi_{i \tau}(z)$ are 
Wannier-like exponentially-localized states, describing the $z$-axis behavior of the TI surface states in the unit cell $i$ and surface $\tau$, 
$\bk = (k_x, k_y)$ is the transverse momentum, and $L_{x,y}$ are the sample dimensions in the $x$- and $y$-directions. 
Due to the exponential localization of the Wannier functions $\phi_{i \tau}(z)$, it is easy to see that, upon substitution of Eq.~\eqref{eq:5} into $H_{int}$, 
the dominant terms will correspond to pairing interaction that is local in both the $i$ and the $\tau$ indices
\beq
\label{eq:6}
H_{int} = - \frac{\tilde U}{L_x L_y} \sum_{\bk \bk' \bq} \sum_{i \tau} c^\dg_{\bk + \bq i \upa \tau} c^\dg_{\bk' - \bq i \da \tau} c^\pdg_{\bk' i \da \tau} c^\pdg_{\bk i \upa \tau}, 
\eeq
where 
\beq
\label{eq:7}
\tilde U = U \int_{-\infty}^{\infty} d z | \phi_{i \tau}(z) |^4. 
\eeq
Redefining $\tilde U d \rightarrow U$ and transforming to the crystal momentum basis with respect to the $i$-index, we then obtain
\beq
\label{eq:8}
H_{int} = - \frac{U}{V} \sum_{\bk \bk' \bq} \sum_{\tau} c^\dg_{\bk + \bq \upa \tau} c^\dg_{\bk' - \bq \da \tau} c^\pdg_{\bk' \da \tau} c^\pdg_{\bk \upa \tau}, 
\eeq
where $\bk$ henceforth means the full 3D crystal momentum and $V = L_x L_y L_z$ is the sample volume. 

As demonstrated in Refs.~\onlinecite{Moore12} and \onlinecite{Aji14}, the electron pairing will predominantly occur in two distinct channels: BCS and FFLO. 
Correspondingly, we simplify $H_{int}$ by leaving only the two contributions, $H^{BCS}_{int}$ and $H^{FFLO}_{int}$, where 
\beq
\label{eq:9}
H^{BCS}_{int} = - \frac{U}{V} \sum_{\bk \bk' \tau}  c^\dg_{\bk \upa \tau} c^\dg_{- \bk  \da \tau} c^\pdg_{- \bk' \da \tau} c^\pdg_{\bk' \upa \tau}, 
\eeq
and 
\beq
\label{eq:10}
H^{FFLO}_{int} = - \frac{U}{V} \sum_{\bk \bk' \bQ \tau}  c^\dg_{\bQ + \bk \upa \tau} c^\dg_{\bQ - \bk  \da \tau} c^\pdg_{\bQ - \bk' \da \tau} c^\pdg_{\bQ + \bk' \upa \tau}, 
\eeq
Here the momentum vector $\bQ$ labels the locations of the Weyl nodes, to be specified below. 
An important property of Eq.~\eqref{eq:10}, which follows from momentum conservation, is that $H^{FFLO}_{int}$ does not couple 
different Weyl nodes. This will play a significant role in selecting the lowest energy superconducting state. 

To proceed we diagonalize the noninteracting part of the Hamiltonian, $H_0$, in order to rewrite the full Hamiltonian in the basis of the 
eigenstates of $H_0$. 
Performing a canonical (i.e. commutation relations preserving) transformation
\beq
\label{eq:11}
\sigma^{\pm} \rightarrow \tau^z \sigma^{\pm}, \,\, \tau^{\pm} \rightarrow \sigma^z \tau^{\pm},
\eeq
which corresponds to the following unitary transformation of the electron creation operators
\beqa
\label{eq:11.5}
&&c^\dg_{\bk \upa T} \rightarrow c^\dg_{\bk \upa T}, \,\,  c^\dg_{\bk \da T} \rightarrow c^\dg_{\bk \da T}, \nonumber \\
&&c^\dg_{\bk \upa B} \rightarrow c^\dg_{\bk \upa B}, \,\, c^\dg_{\bk \da B} \rightarrow - c^\dg_{\bk \da B},
\eeqa
one obtains
\beq
\label{eq:12}
H_0 = v _F (\hat z \times \bsigma) \cdot \bk + \hat m(k_z) \sigma^z, 
\eeq
where $\hat m(k_z) = b + \hat t(k_z)$. 
The pairing terms are clearly unchanged by this transformation. 

Diagonalizing the $2 \times 2$ matrix $\hat m(k_z)$ one obtains the following eigenvalues 
\beq
\label{eq:13}
m_r(k_z) = b +  r t(k_z), 
\eeq
where $r = \pm$ and $t(k_z) = \sqrt{t_S^2 + t_D^2 + 2 t_S t_D \cos(k_z d)}$. 
The corresponding eigenvectors are 
\beq
\label{eq:14}
|u^r(k_z) \rangle = \frac{1}{\sqrt{2}} \left(1, r \frac{t_S + t_D e^{- i k_z d}}{t(k_z)} \right). 
\eeq
Diagonalizing the remaining $2 \times 2$ blocks of $H_0$ one finally obtains its eigenvalues
\beq
\label{eq:15}
\epsilon_{s r}(\bk) = s \epsilon_r(\bk) = s \sqrt{v_F^2 (k_x^2 + k_y^2) + m_r^2(k_z)}, 
\eeq
where $s = \pm$ and the corresponding eigenvectors
\beq
\label{eq:16} 
|v^{s r}(k_z) \rangle = \frac{1}{\sqrt{2}} \left(\sqrt{1 + s \frac{m_r(k_z)}{\epsilon_r(k_z)}}, - i s e^{i \varphi} \sqrt{1 - s \frac{m_r(k_z)}{\epsilon_r(k_z)}} \right), 
\eeq
where $e^{i \varphi} = \frac{k_x + i k_y}{\sqrt{k_x^2 + k_y^2}}$. 
The full 4-component eigenvector may be viewed as a tensor product
\beq
\label{eq:17}
|z^{s r}(k_z) \rangle = |u^r(k_z) \rangle \otimes |v^{s r}(k_z) \rangle. 
\eeq
The Weyl nodes correspond to points along the $z$-axis in momentum space at which $m_-(k_z) = b - t(k_z)$ vanishes. 
Solving the equation $b = t(k_z)$, one obtains $\bQ = Q \hat z$, where 
\beq
\label{eq:18}
Q = \frac{\pi}{d} \pm \frac{1}{d} \arccos\left(\frac{t_S^2 + t_D^2 - b^2}{2 t_S t_D} \right) \equiv \frac{\pi}{d} \pm k_0. 
\eeq

When the Fermi energy is sufficiently close to the Weyl nodes, namely when $\epsilon_F \ll b$, the Fermi level only crosses the $s = +, r = -$ band, assuming 
$\epsilon_F > 0$ for concreteness. 
Projecting onto this band, we write the electron creation operators as
\beq
\label{eq:19}
c^\dg_{\bk \sigma \tau} = z^{* + -}_{\sigma \tau}(\bk) c^\dg_{\bk + -} \equiv z^*_{\sigma \tau}(\bk) c^\dg_\bk,
\eeq
i.e. we will omit the explicit $s = +,\, r= -$ indices from now on in all the equations for brevity. 
Substituting this into $H^{BCS}_{int}$ and $H^{FFLO}_{int}$, we finally obtain the following projected low energy BCS and FFLO Hamiltonians
\beq
\label{eq:20}
H_{BCS} = \sum_\bk \xi(\bk) c^\dg_\bk c^\pdg_\bk - \frac{U}{2 V} \sum_{\bk \bk'} f^*_\bk f_{\bk'} c^\dg_\bk c^\dg_{- \bk} c^\pdg_{-\bk'} c^\pdg_{\bk'}, 
\eeq
and 
\beqa
\label{eq:21}
&&H_{FFLO} = \sum_\bk \xi(\bk) c^\dg_\bk c^\pdg_\bk \nonumber \\
&-&\frac{U}{2 V} \sum_{\bk \bk' \bQ} \tilde f^*_{\bk \bQ} \tilde f_{\bk' \bQ} c^\dg_{\bQ + \bk} c^\dg_{\bQ - \bk} c^\pdg_{\bQ -\bk'} c^\pdg_{\bQ + \bk'},
\eeqa
where in Eq.~\eqref{eq:21} we assume that $Q d \ll 1$, i.e. Weyl nodes are far away from the BZ boundary. 
$\xi(\bk) = \epsilon(\bk) - \epsilon_F$ is the band energy, counted from the Fermi energy and 
\beq
\label{eq:22}
f_\bk = \frac{i e^{i \varphi}}{2} \sqrt{1 - \frac{m^2(k_z)}{\epsilon^2(\bk)}}, 
\eeq
while 
\beqa
\label{eq:23}
\tilde f_{\bk \bQ}&= &\frac{i e^{i \varphi}}{4} \left[ \sqrt{1 + \frac{m(Q + k_z)}{\epsilon(\bQ + \bk)}} \sqrt{1 - \frac{m(Q - k_z)}{\epsilon(\bQ - \bk)}}\right. \nonumber \\
&+& \left. \sqrt{1 + \frac{m(Q - k_z)}{\epsilon(\bQ - \bk)}} \sqrt{1 - \frac{m(Q + k_z)}{\epsilon(\bQ + \bk)}} \right]. 
\eeqa
The nontrivial momentum-dependent form-factors $f_\bk$ and $\tilde f_{\bk \bQ}$ are a consequence of projection of the pairing interaction terms onto a nondegenerate band. 
Their properties play an important role in the physics of the BCS and the FFLO states.
It is easy to see that the BCS form-factor vanishes identically everywhere on the $k_x = k_y = 0$ line in momentum space. 
The four points, at which this line intersects the two Fermi surface sheets will produce four nodes in the BCS gap function, as will be seen below. 
On the other hand, the FFLO form-factor $\tilde f_{\bk \bQ}$ never vanishes, which is directly related to the fact that the function $m(k_z)$ changes sign at the 
Weyl node locations $k_x = k_y = 0, k_z =Q$. This means that the FFLO state is nodeless. 
It would then seem natural if the fully-gapped FFLO state would have a larger condensation energy and thus a lower overall energy than the nodal BCS state, with the 
gap vanishing at four points in the BZ. 
Surprisingly, as we show below, this is not the case: the BCS state in fact has a lower energy. 

\section{Condensation energies of the BCS and the FFLO states}
\label{sec:3}
In this section we will evaluate and compare the condensation energies of the BCS and the FFLO states, using the low-energy Hamiltonians, derived in the previous section. 
\subsection{BCS state}
\label{sec:3.1}
We analyze $H_{BCS}$ using the standard mean-field theory. 
The mean-field BCS Hamiltonian has the form
\beq
\label{eq:24}
H_{BCS} = \sum_\bk \left[\xi(\bk) c^\dg_\bk c^\pdg_\bk - \frac{\Delta}{2} ( f^*_\bk c^\dg_\bk c^\dg_{-\bk} + f_\bk c^\pdg_{-\bk} c^\pdg_\bk) \right] + \frac{V}{2 U} \Delta^2, 
\eeq
where the pairing amplitude $\Delta$ is given by
\beq
\label{eq:25}
\Delta = \frac{U}{V} \sum_\bk f^*_\bk \langle c^\dg_\bk c^\dg_{-\bk} \rangle = \frac{U}{V} \sum_\bk f_\bk \langle c^\pdg_{-\bk} c^\pdg_{\bk} \rangle.
\eeq 
It is important to note that our BCS state has {\em odd parity} since $f_\bk$ changes sign under inversion. This is already in contrast to Ref.~\onlinecite{Moore12}, which 
claimed an even-parity BCS state. 

Diagonalizing $H_{BCS}$ using Bogoliubov transformation, one obtains
\beq
\label{eq:26}
H_{BCS} = \sum_\bk E(\bk) \psi^\dg_\bk \psi^\pdg_\bk + \frac{1}{2} \sum_\bk [\xi(\bk) - E(\bk)] + \frac{V }{2 U} \Delta^2. 
\eeq
Here $\psi^\dg_\bk$ are the Bogoliubov quasiparticle creation operators, 
\beq
\label{eq:27}
E(\bk) = \sqrt{\xi^2(\bk) + |f_{\bk}|^2 \Delta^2}, 
\eeq
is the quasiparticle energy and the pairing amplitude $\Delta$ satisfies the standard BCS equation (assuming temperature $T = 0$)
\beq
\label{eq:28}
1 = \frac{U}{2 V} \sum_\bk \frac{|f_\bk|^2}{E(\bk)}.
\eeq
It is clear from Eq.~\eqref{eq:27} that the quasiparticle energy indeed vanishes at four points in the first BZ, at which the form-factor $f_\bk$ vanishes. 

The $T = 0$ BCS equation may be easily solved following the well-known steps.~\cite{deGennes}
One obtains
\beq
\label{eq:29}
\Delta = 2 \omega_D e^{- \langle |f_\bk|^2 \ln |f_\bk| \rangle/ \langle |f_\bk|^2 \rangle} e^{- 1/ U g(\epsilon_F) \langle |f_\bk|^2 \rangle}. 
\eeq
Here $\omega_D$ is the Debye frequency,
\beqa
\label{eq:30}
g(\epsilon_F)&=&\int \frac{d^3 k}{(2 \pi)^3} \delta[\epsilon(\bk) - \epsilon_F] \nonumber \\
&=&\frac{\epsilon_F}{4 \pi^2 v_F^2} \int_{-\pi/d}^{\pi/d} d k_z \Theta[\epsilon_F - |m(k_z)|], 
\eeqa
 is the density of states at Fermi energy, and 
\beq
\label{eq:31}
\langle |f_\bk|^2 \rangle = \frac{1}{g(\epsilon_F)} \int \frac{d^3 k}{(2 \pi)^3} |f_\bk|^2 \delta[\epsilon(\bk) - \epsilon_F], 
\eeq
is the Fermi surface average of the BCS gap function $|f_\bk|^2$. 

In the limit $\epsilon_F \ll b$, i.e. when the Fermi energy is close to the Weyl nodes, the Fermi surface average may be easily evaluated explicitly. 
Indeed, in this case the band dispersion in the $z$-direction in momentum space may be assumed to be linear, which follows from the 
leading-order expansion of $m(k_z)$ near the nodes
\beq
\label{eq:32}
m(k_z) \approx m(Q) + \left.\frac{d m(k_z)}{d k_z} \right|_{k_z = Q} \delta k_z = \pm \tilde v_F \delta k_z, 
\eeq
where 
\beq
\label{eq:33}
\tilde v_F = \frac{d}{2 b} \sqrt{[b^2 - (t_S - t_D)^2][(t_S + t_D)^2 - b^2]}, 
\eeq
is the $z$-component of the Fermi velocity near the nodes. 
Then one obtains
\beq
\label{eq:34}
g(\epsilon_F) = \frac{\epsilon_F^2}{\pi^2 v_F^2 \tilde v_F}, 
\eeq
and 
\beqa
\label{eq:35}
\langle |f_\bk|^2 \rangle&=&\frac{1}{4} \left[1 - \frac{\langle m^2(k_z) \rangle}{\epsilon_F^2} \right] \nonumber \\
&=&\frac{1}{4}\left[1 - \frac{\tilde v_F^3}{2 \epsilon_F^3} \int^{\epsilon_F/\tilde v_F}_{-\epsilon_F /\tilde v_F}\,  d k_z \, k_z^2\right] = \frac{1}{6}. 
\eeqa
The pairing amplitude is thus given by
\beq
\label{eq:36}
\Delta \approx  5 \omega_D e^{-6/U g(\epsilon_F)}. 
\eeq

To evaluate the condensation energy we take the expectation value of $H_{BCS}$ at $T = 0$.
In this case there are no quasiparticles and we obtain
\beq
\label{eq:37}
\frac{E_{BCS}}{V} = \frac{1}{2 V} \sum_\bk [\xi(\bk) - E(\bk)] + \frac{\Delta^2}{2 U}. 
\eeq
This is again evaluated in the standard way~\cite{deGennes} and gives the following result for the condensation energy, i.e. the energy gain 
compared to the normal state energy $E_0$
\beqa
\label{eq:38}
\frac{E_{BCS} - E_0}{V}&=&- \frac{1}{4} g(\epsilon_F) \langle |f_\bk|^2 \rangle \Delta^2 \nonumber \\
&\approx&- \frac{25 \omega_D^2 g(\epsilon_F)}{24} e^{- 12/U g(\epsilon_F)}. 
\eeqa

\subsection{FFLO state}
\label{sec:3.2}
The FFLO state is analyzed in exactly the same way as above.
One important point to note is that the FFLO Hamiltonian Eq.~\eqref{eq:21} clearly does not mix different Weyl nodes, i.e. different values of $\bQ$, as 
required by momentum conservation. This means that, in mean-field theory, the contributions of different Weyl nodes may be analyzed separately
and simply summed when calculating the total condensation energy. 

The mean-field Hamiltonian for a single Weyl node, i.e. with a fixed $\bQ$, is given by
\beqa
\label{eq:39}
H^{\bQ}_{FFLO}&=&\sum_\bk \left[\xi(\bk) c^\dg_\bk c^\pdg_\bk - \frac{\Delta}{2} \left(\tilde f^*_{\bk \bQ} c^\dg_{\bQ + \bk} c^\dg_{\bQ - \bk}\right. \right. \nonumber \\ 
&+&\left. \left. \tilde f_{\bk \bQ} c^\pdg_{\bQ - \bk} c^\pdg_{\bQ + \bk}\right) \right] + \frac{V}{2 U} \Delta^2, 
\eeqa
where the pairing amplitude is
\beq
\label{eq:40}
\Delta = \frac{U}{V} \sum_\bk \tilde f^*_{\bk \bQ} \langle c^\dg_{\bQ + \bk} c^\dg_{\bQ -\bk} \rangle = \frac{U}{V} \sum_\bk \tilde f_{\bk \bQ} \langle
 c^\pdg_{\bQ - \bk} c^\pdg_{\bQ + \bk} \rangle.
\eeq 
Diagonalizing Eq.~\eqref{eq:39} one obtains
\beqa
\label{eq:41}
H^\bQ_{FFLO}&=&\sum_\bk \left[\frac{\epsilon(\bQ + \bk) - \epsilon(\bQ - \bk)}{2} + E(\bk) \right] \psi^\dg_\bk \psi^\pdg_\bk \nonumber \\
&+&\frac{1}{2} \sum_\bk [\xi(\bk) - E(\bk)] + \frac{V}{2 U} \Delta^2,
\eeqa
where 
\beq
\label{eq:42}
E(\bk) = \sqrt{\left[\frac{\xi(\bQ + \bk) + \xi(\bQ - \bk)}{2} \right]^2 + |\tilde f_{\bk \bQ}|^2 \Delta^2}. 
\eeq
A crucial difference between the FFLO and the BCS states is the term $[\epsilon(\bQ + \bk) - \epsilon(\bQ - \bk)]/2$ in the energy of the FFLO Bogoliubov 
quasiparticles in Eq.~\eqref{eq:41}. 
This term is a consequence of the fact that the states with momenta $\bQ + \bk$ and $\bQ - \bk$ are not related by any exact symmetry, unlike $\bk$ and 
$-\bk$ in the BCS case, which are related by inversion.  
An important property of a Weyl metal, however, which may allow the FFLO state to exist in principle, is the low-energy chiral symmetry, which emerges 
as $\epsilon_F \rightarrow 0$ and which implies that 
\beq
\label{eq:43}
\epsilon(\bQ + \bk)  \approx \epsilon(\bQ - \bk), 
\eeq
the equality becoming more and more precise as the energy is lowered. 
Let us first assume that Eq.~\eqref{eq:43} holds exactly, as it would in a low-energy model of a Weyl metal with an exactly 
linear dispersion. 
Then we obtain
\beq
\label{eq:44}
H^\bQ_{FFLO} = \sum_\bk E(\bk) \psi^\dg_\bk \psi^\pdg_\bk + \frac{1}{2} \sum_\bk [\xi(\bk) - E(\bk)] + \frac{V}{2 U} \Delta^2. 
\eeq
The mean-field equation for the pairing amplitude, again assuming Eq.~\eqref{eq:43} holds, takes the form, identical to the BCS case
\beq
\label{eq:45}
1 = \frac{U}{2 V} \sum_\bk \frac{|\tilde f_{\bk \bQ}|^2}{E(\bk)}.
\eeq

To proceed, we now evaluate explicitly the FFLO gap function. After elementary algebra, we obtain
\beqa
\label{eq:46}
|\tilde f_{\bk \bQ}|^2&=&\frac{1}{8}\left[1 - \frac{m(Q + k_z) m(Q - k_z)}{\epsilon_F^2} \right. \nonumber \\
&+&\left. \sqrt{1 - \frac{m^2(Q + k_z)}{\epsilon_F^2}} \sqrt{1 - \frac{m^2(Q - k_z)}{\epsilon_F^2}} \right]. 
\eeqa
Low energy chiral symmetry implies that 
\beq
\label{eq:47}
m(Q + k_z ) \approx - m(Q - k_z). 
\eeq
Assuming this to hold, we finally obtain
\beq
\label{eq:48}
|\tilde f_{\bk \bQ} |^2 = \frac{1}{4}, 
\eeq 
which is by a factor of $3/2$ larger than the corresponding result for the Fermi surface average of the BCS gap function, Eq.~\eqref{eq:35}. 
This is to be expected, since the FFLO gap function, unlike the BCS one, does not have nodes.
This result might then suggest that the FFLO state should have a lower energy. 
However, this turns out not to be the case. The culprit is another factor that influences the magnitude of the pairing amplitude, i.e. the 
density of states. 
Since in mean-field theory of the FFLO state the two Weyl nodes are completely decoupled from each other, only half of the total density 
of states determines the magnitude of the gap for each Fermi surface sheet. 
We obtain
\beq
\label{eq:49}
\Delta = 4 \omega_D e^{- 2/ U g(\epsilon_F) \langle |\tilde f_{\bk \bQ}|^2 \rangle} = 4 \omega_D e^{- 8/ U g(\epsilon_F)}.
\eeq
The extra factor of 2 in the expression above, compared to the corresponding Eq.~\eqref{eq:29} in the BCS case, arises precisely from the fact that 
the density of states per Weyl node is $g(\epsilon_F)/2$, $g(\epsilon_F)$ being the total density of states. 
It is then clear that the magnitude of $\Delta$ in the FFLO state is exponentially smaller than in the BCS state in the weak coupling regime $U g(\epsilon_F) \ll 1$.

Summing the identical contributions from the two Weyl nodes, the total FFLO condensation energy is finally given by
\beq
\label{eq:50}
\frac{E_{FFLO} - E_0}{V} = - \frac{1}{16} g(\epsilon_F) \Delta^2 = - \omega_D^2 g(\epsilon_F) e^{- 16/ U g(\epsilon_F)}. 
\eeq
Comparing to the corresponding result in the BCS case, Eq.~\eqref{eq:38}, it is clear that the FFLO state condensation energy 
is exponentially smaller in the weak coupling regime, within our model of a Weyl metal. 
Since the critical temperature $T_c \sim \Delta$ at weak coupling, this also implies that $T_c$ of the BCS state is higher than $T_c$ of the FFLO state and 
we thus do not expect any transitions between them as a function of temperature. 

In reality the situation is even worse for the FFLO state, however, since we have so far been assuming exact chiral symmetry, expressed by Eq.~\eqref{eq:43}. 
But chiral symmetry is not exact, and it is instructive to work out the consequences of that. 
While we have already shown that, in our model, the FFLO state is not favored even when the chiral symmetry exists, this result is based on comparing energies, and 
thus may not be universal. The arguments presented below are of a more general validity. 
  
Let us expand $\epsilon(\bQ \pm \bk)$ in Taylor series with respect to the deviation from the Weyl node location $\bk$, assuming $\bk$ is small. 
Let us also take $\bk = k_z \hat z$ for the sake of simplicity. 
We obtain
\beqa
\label{eq:51}
&&\epsilon(Q + k_z) - \epsilon(Q - k_z) \approx \left. \frac{d \epsilon}{d k_z} \right|_{k_z = Q +} k_z + \left. \frac{d \epsilon}{d k_z} \right|_{k_z = Q -} k_z \nonumber \\
&+&\frac{1}{2} \left. \frac{d^2 \epsilon}{d k_z^2} \right|_{k_z = Q +} k_z^2 - \frac{1}{2} \left. \frac{d^2 \epsilon}{d k_z^2} \right|_{k_z = Q -} k_z^2 + \ldots,
\eeqa
where we have taken into account the fact that the band velocity $d \epsilon/ d k_z$ is discontinuous and changes sign at the Weyl node locations. 
Since 
\beq
\label{eq:52}
\left. \frac{d \epsilon}{d k_z} \right|_{k_z = Q +} = -  \left. \frac{d \epsilon}{d k_z} \right|_{k_z = Q -}, 
\eeq
the first order term in the expansion above vanishes, which is precisely the expression of the approximate chiral symmetry. 
The quadratic term, however, does not vanish and is given by
\beqa
\label{eq:53}
&& \frac{1}{2} \left. \frac{d^2 \epsilon}{d k_z^2} \right|_{k_z = Q +} - \frac{1}{2} \left. \frac{d^2 \epsilon}{d k_z^2} \right|_{k_z = Q -} = 
\left. \frac{d^2 m(k_z)}{d k_z^2} \right|_{k_z = Q} \nonumber \\
&=& d^2 \frac{b^4 - (t_S^2 - t_D^2)^2}{4 b^3} \sim d^2 b. 
\eeqa
We may expect the FFLO state to exist as a local minimum of the free energy as long as
\beq
\label{eq:54}
\max_{\bk} |\epsilon(\bQ + \bk) - \epsilon(\bQ - \bk)| \lesssim \Delta, 
\eeq
where the maximum is taken over states on the Fermi surface.  
Taking $k_z \sim \epsilon_F/ \tilde v_F$, where $\tilde v_F \sim d \, t_S$ is the Fermi velocity at the Weyl nodes, and assuming $b \gg |t_S - t_D|$, 
Eq.~\eqref{eq:54} becomes
\beq
\label{eq:55}
b \,\epsilon_F^2/ t_S^2 < \Delta, 
\eeq
where $t_S$ should be regarded as the highest energy scale in the problem, of the order of the total bandwidth.  
Using Eqs.~\eqref{eq:34} and \eqref{eq:49}, we may rewrite this inequality as
\beq
\label{eq:56}
\frac{b}{\omega_D} \frac{\epsilon_F^2}{t_S^2} \lesssim \exp\left( - \frac{16 \pi^2 v_F^2 \tilde v_F}{U \epsilon_F^2}\right), 
\eeq
Using $v_F \sim \tilde v_F \sim t_S d$, it becomes clear that this inequality is very hard, if not impossible, to satisfy for any reasonable values of the relevant parameters. 

The above arguments lead us to the conclusion that, when the inversion symmetry is present and when the coupling is weak, nodal odd-parity BCS superconducting state, which is a close analog of the A phase in $^3$He, is much more likely to be realized in a Weyl metal. 
This state is topologically nontrivial and is characterized by an interesting edge state structure, in which Fermi arcs of the ``parent" nonsuperconducting Weyl semimetal coexist 
with Majorana edge states of the nodal BCS superconductor, as will be described in detail in the following section. 

\section{Majorana and Fermi arc edge modes in the nodal BCS state}
\label{sec:4}
In this section we will discuss in some detail the nontrivial momentum-space topology of the nodal BCS state and the corresponding Majorana and Fermi arc
edge states. 
Some work on this has already been done before in Refs.~\onlinecite{Meng12,Tanaka15}, and the discussion below mostly serves to provide a clear connection between these 
previously published results and place them in the context of our model. 

To discuss momentum-space topology, which involves not only states near the Fermi energy, it will be necessary to consider the BCS Hamiltonian without making the projection 
onto the low-energy states only. We will still restrict ourselves to the $r = -$ states, since these are the bands that touch at the Weyl nodes in the nonsuperconducting state,
but will take into account both bands that touch. The mean-field BCS Hamiltonian then takes the form
\beqa
\label{eq:57}
H&=&\sum_\bk [v_F (\hat z \times \bsigma) \cdot \bk + m(k_z) \sigma^z - \epsilon_F] c^\dg_\bk c^\pdg_\bk \nonumber \\
&-& \frac{1}{2} \sum_\bk \left(\Delta c^\dg_{\bk \upa} c^\dg_{- \bk \da}  + \Delta^* c^\pdg_{- \bk \da} c^\pdg_{\bk \upa} \right), 
\eeqa
where $m(k_z) \equiv m_-(k_z)$,
\beq
\label{eq:58}
\Delta = \frac{U}{V} \sum_\bk \langle c^\pdg_{- \bk \da} c^\pdg_{\bk \upa} \rangle, 
\eeq
and spin indices were suppressed in the first line of Eq.~\eqref{eq:57} for brevity. 
To analyze Eq.~\eqref{eq:57} we introduce a Nambu spinor
\beq
\label{eq:59}
\psi_\bk = (c^\pdg_{\bk \upa}, c^\pdg_{\bk \da}, c^\dg_{- \bk \da}, c^\dg_{- \bk \upa}) \equiv (\psi_{\bk 1 \upa}, \psi_{\bk 1 \da}, \psi_{\bk 2 \upa}, \psi_{\bk 2 \da}). 
\eeq
In the Nambu spinor notation, the BCS Hamiltonian takes the following form
\beqa
\label{eq:60}
H&=&\frac{1}{2} \sum_\bk [v_F (\hat z \times \bsigma) \cdot \bk + m(k_z) \sigma^z - \epsilon_F \kappa^z] \psi^\dg_\bk \psi^\pdg_\bk \nonumber \\
&-&\frac{1}{8} \sum_\bk (\Delta \kappa^+ + \Delta^* \kappa^-) \sigma^z \psi^\dg_\bk \psi^\pdg_\bk,
\eeqa
where $\bkappa$ is the Nambu pseudospin. 
\begin{figure}[t]
  \includegraphics[width=8cm]{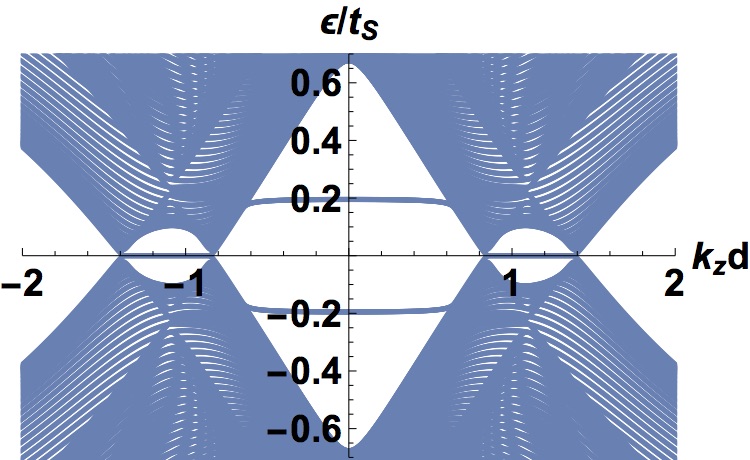}
  \caption{(Color online) Numerically calculated eigenstate dispersions for Eq.~\eqref{eq:60} along the $z$-direction in momentum space for a sample of finite width in the $x$-direction. 
  Parameters in Eq.~\eqref{eq:60} are chosen as follows: $t_S = 1, t_D = 0.9, b =1, \epsilon_F = \Delta = 0.2$. 
  Majorana modes are pinned to zero energy and exist in intervals of width of order $|\Delta|/\tilde v_F$ around each Weyl node. Fermi arc edge modes are
  split into particle-hole antisymmetric and symmetric branches at $\pm \epsilon_F$.}  
  \label{fig:1}
\end{figure} 

It is instructive to start from the limit of $\epsilon_F = 0$, in which case Eq.~\eqref{eq:60} reduces to the model of Ref.~\onlinecite{Meng12}. 
In this case it is clear that the Nambu pseudospin block of the Hamiltonian may be diagonalized separately and we obtain
\beq
\label{eq:61}
H = \frac{1}{2} \sum_\bk \left\{v_F (\hat z \times \sigma) \cdot \bk + [m(k_z) + p |\Delta|/2] \sigma^z \right\} \psi^\dg_\bk \psi^\pdg_\bk, 
\eeq
where $p = \pm$ labels the two eigenvalues $\pm |\Delta|$ of the matrix $(\Delta \kappa^+ + \Delta^* \kappa^-)/2$. 
Apart from the factor of $1/2$ in front, which expresses the doubling of degrees of freedom in the Nambu pseudospin notation, Eq.~\eqref{eq:61} may be viewed 
as the Hamiltonian of two massive 2D Dirac fermions with the masses $m(k_z) \pm |\Delta|/2$ varying as functions of parameter $k_z$. 
Sign change of the Dirac masses signals quantum Hall transitions. 
In particular, when $\Delta = 0$, $m(k_z)$ changes sign at the locations of the two Weyl nodes, given by $\pi/d \pm k_0$. 
The topologically nontrivial range of $k_z$, where $m(k_z) > 0$, corresponds to the range of momenta in which chiral Fermi arc edge states exist on 
any sample surface, not perpendicular to the $z$-axis.~\cite{Burkov11-1}
Due to doubling of the number of degrees of freedom in Eq.~\eqref{eq:61}, these chiral Fermi arc edge states come in degenerate particle-hole symmetric and 
antisymmetric pairs. 
Turning on a nonzero $\Delta$, the Dirac mass of the particle-hole symmetric states decreases by $|\Delta|/2$ while the mass of the particle-hole antisymmetric states increases by the same amount. There is thus an interval of momenta near each Weyl node location of width 
\beqa
\label{eq:62}
\delta k_z&=&\frac{1}{d}\left[\arccos\left(\frac{t_S^2 + t_D^2 - (b + |\Delta|/2)^2}{2 t_S t_D}\right)\right. \nonumber \\
&-&\left.\arccos\left(\frac{t_S^2 + t_D^2 - (b - |\Delta|/2)^2}{2 t_S t_D}\right)\right] \approx \frac{|\Delta|}{\tilde v_F}, 
\eeqa
in which only the particle-hole antisymmetric states are topologically nontrivial (the last equality holds assuming $|\Delta| \ll b$). 
These momentum intervals give rise to chiral Majorana edge modes, which do not have particle-hole symmetric partners, unlike the ``ordinary" Fermi arc edge states. 
\begin{figure}[t]
\subfigure{
\label{fig:2a}
  \includegraphics[width=8cm]{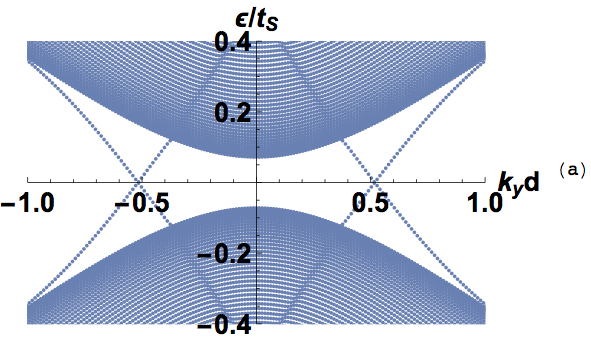}}
  \subfigure{
  \label{fig:2b}
  \includegraphics[width=8cm]{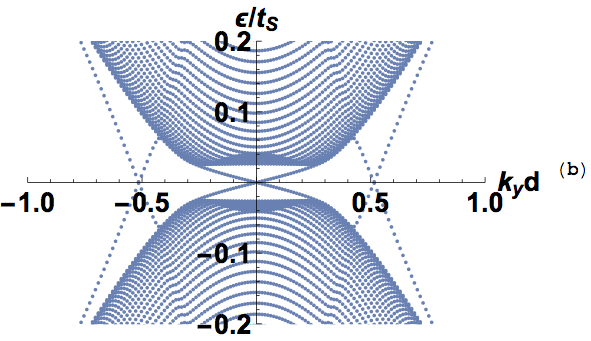}}
  \subfigure{
  \label{fig:2c}
  \includegraphics[width=8cm]{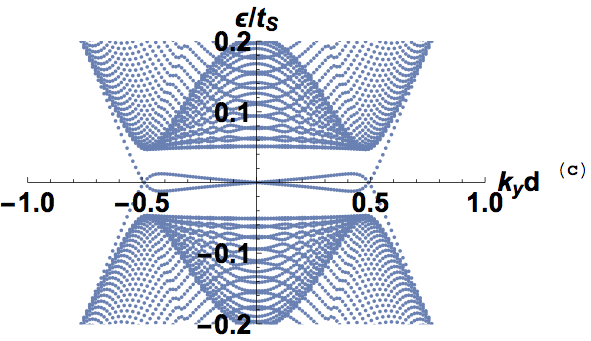}}
  \caption{(Color online) Numerically calculated eigenstate dispersions for Eq.~\eqref{eq:60} along the $y$-direction in momentum space for a sample of finite width in the $x$-direction, 
  at several values of $k_z$. 
  Parameters in Eq.~\eqref{eq:61} are chosen as follows: $t_S = 1, t_D = 0.9, b =1, \epsilon_F = 0.5, \Delta = 0.1$.
  (a) $k_z$ is outside the Fermi surface. Two pairs (corresponding to two surfaces) of chiral Fermi arc modes, crossing at $\pm \epsilon_F$ at $k_y = 0$ are visible.
  (b) $k_z$ is just inside the Fermi surface. Two chiral Majorana modes are visible now, crossing at $k_y = 0$ at zero energy.
  (c) $k_z$ is closer to a Weyl node location. Majorana modes directly connect with the Fermi arcs.}  
  \label{fig:2}
\end{figure} 
The Majorana edge states terminate at points at which $m(k_z) = \pm |\Delta|/2$. These are point nodes in the spectrum of the Bogoliubov quasiparticles. 
Each Weyl node in the normal state thus splits into two Bogoliubov-Weyl nodes in the superconducting state, which inherit the chirality of the parent Weyl node, as seen from 
Eq.~\eqref{eq:61}. 

When $\epsilon_F > 0$, Eq.~\eqref{eq:60} no longer has the simple form of two independent massive Dirac Hamiltonians, but the eigenstate spectrum is still easily found. 
We obtain
\beq
\label{eq:63}
E^2_{\pm}(\bk) = \epsilon^2(\bk) + \frac{|\Delta|^2}{4} + \epsilon_F^2 \pm \sqrt{4\epsilon^2(\bk) \epsilon_F^2 + |\Delta|^2 m^2(k_z)}. 
\eeq
The node locations are now given by the solutions of the equation
\beq
\label{eq:64}
|m(k_z)| = \sqrt{\epsilon_F^2 + \frac{|\Delta|^2}{4}}. 
\eeq
The edge state spectrum may be 
easily found numerically, as shown in Fig.~\ref{fig:1}. 
The degeneracy of the Fermi arc doublet is split, with the particle-hole symmetric branch moving down to $-\epsilon_F$ in energy, while the particle-hole antisymmetric 
branch moves up to $\epsilon_F$. 
The Majorana states, which extend between the nodes, remain pinned at zero energy (note that zero is at the Fermi energy here). 

In the limit $|\Delta| \ll \epsilon_F$, we may obtain a simple picture of the Majorana states from the BCS Hamiltonian, projected onto the low-energy $s = +$ states
\beq
\label{eq:65}
H = \frac{1}{2} \sum_\bk\left[ \xi(\bk) \kappa^z - \frac{1}{2} (\Delta f_\bk^* \kappa^+ + \Delta^* f_\bk \kappa^-)\right] \psi^\dg_\bk \psi^\pdg_\bk, 
\eeq
where $\psi_\bk = (c^\pdg_\bk, c^\dg_{-\bk})$ is the projected Nambu spinor and $\xi(\bk) = \sqrt{v_F^2(k_x^2 + k_y^2) + m^2(k_z)} - \epsilon_F$. 
This again has the form of the Hamiltonian of a 2D Dirac fermion, with the mass $\xi(0,0,k_z)$, which depends on $k_z$ as a parameter. 
The mass changes sign at points, satisfying the equation $|m(k_z)| = \epsilon_F$, which coincides with the locations of the nodes of the 
superconducting gap function $f_\bk$. 
The topologically nontrivial momentum range, which in this case corresponds to negative Dirac mass, coincides with the range of momenta inside the two 
Fermi surface sheets, enclosing the Weyl nodes.
There are thus Majorana zero-energy edge states, which exist on arcs, connecting the gap function nodes of the opposite sides of each Fermi surface sheet. 
For sample surfaces perpendicular, say, to the $x$-direction, the dispersion of the Majorana edge modes in the $y$-direction is chiral and exist
for a given $k_z$ within the energy interval $\pm |f_\bk \Delta|$. 
Near the center of each Fermi surface sheet, i.e. close to the locations of the Weyl nodes along the $z$ axis, the superconducting gap is largest in magnitude and the 
Majorana edge state dispersion extends all the way to the Fermi surface, where it gets reconnected with the Fermi arc states, as shown in Fig.~\ref{fig:2}. 
\begin{figure}[t]
\subfigure{
\label{fig:3a}
  \includegraphics[width=8cm]{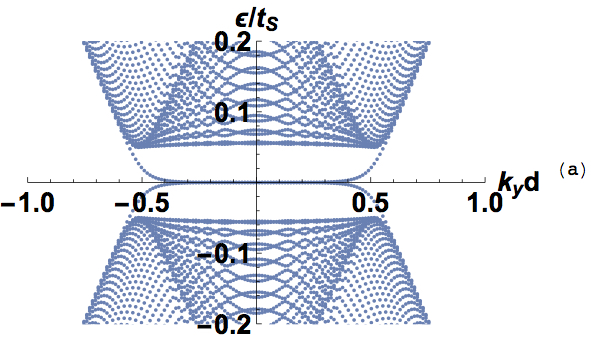}}
  \subfigure{
  \label{fig:3b}
  \includegraphics[width=8cm]{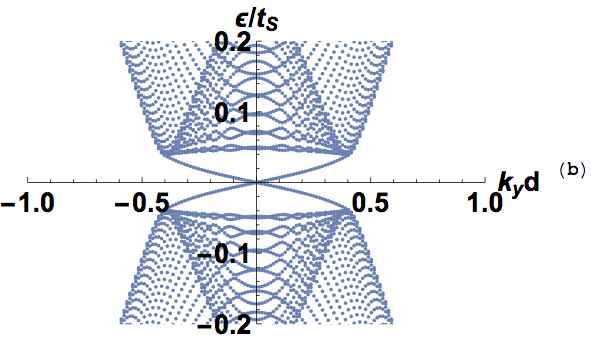}}
   \caption{Same parameters as in Fig.~\ref{fig:2}, but values of $k_z$ are taken near (a) and beyond (b) the Weyl node location.
   Disapperance of the Fermi arcs necessitates a ``flat band" at the transition point. The chirality of the Majorana modes changes sign at the 
   transition. The transition itself occurs exactly at the Weyl node location in the limit $\Delta/\epsilon_F \rightarrow 0$, but is shifted slightly away for any 
   finite $\Delta$.}
    \label{fig:3}
\end{figure} 

However, there is a subtlety here, since the projected low-energy Hamiltonian Eq.~\eqref{eq:65} does not actually give the correct Majorana edge mode dispersion in the $y$-direction. 
The reason is that the two Bogoliubov-Weyl nodes on each Fermi surface sheet have the same chirality, inherited from the Weyl node, enclosed by the Fermi surface sheet, as 
discussed above. 
This can not be deduced from Eq.~\eqref{eq:65} and requires analysis of the full unprojected Hamiltonian Eq.~\eqref{eq:60}. 
Another way to see this is to notice that the Majorana modes must connect with the Fermi arcs, which are not low-energy modes and require the full Hamiltonian Eq.~\eqref{eq:60} to describe. 
An interesting consequence of this, first described in Ref.~\onlinecite{Tanaka15}, is a topological transition that happens when one crosses the Weyl node locations along the $z$-axis.
The Fermi arcs must disappear on approaching Weyl nodes, which means that the finite-$k_z$ crossing points in the edge state dispersions in Fig.~\ref{fig:2} must disappear. 
This leads unavoidably to the appearance of a dispersionless ``flat band" at the value of $k_z$ at which the Fermi arcs disappear, as shown in Fig.~\ref{fig:3a}. 
As $k_z$ is further increased, the Majorana dispersion again becomes chiral, but with chirality of opposite sign, as seen in Fig.~\ref{fig:3b}. 

\section{Discussion and conclusions}
\label{sec:5}
We will start this section by discussing differences between our results and the ones obtained previously in Refs.~\onlinecite{Moore12,Aji14}, which both claimed that the FFLO state has a lower energy. 
In Ref.~\onlinecite{Aji14} a specific low-energy linear-dispersion model for a Weyl metal was used, in which the BCS interaction, Eq.~\eqref{eq:9} vanished identically. 
This is certainly possible in some realizations of Weyl metals, but can not be a general feature. 
Moreover, such an exact cancellation would presumably not happen even within the model of Ref.~\onlinecite{Aji14}, if chiral symmetry violating corrections to the low energy model, which are always present, as discussed above, were included. 
However, in agreement with our results, all superconducting states, found in Ref.~\onlinecite{Aji14}, had odd parity. 

Ref.~\onlinecite{Moore12} considered a model, quite similar to ours as far as its symmetry properties are concerned, also assuming 
a strictly linear band dispersion in all calculations. 
However, the BCS state, found in this reference, was claimed to have even parity, which disagrees with our results. 
The FFLO state in this reference also appears to be different, with a single order parameter for both nodes, while in our case there are two independent
(in mean-field theory) order parameters. In fact, a single order parameter would be impossible in our case: this would imply internode pair scattering
already at the level of mean-field theory, which violates momentum conservation, as clearly seen from Eq.~\eqref{eq:10}. 

We will conclude by pointing out one possible situation in which the FFLO, rather than the nodal BCS state, may be realized. 
FFLO state may happen to be the ground state if the inversion symmetry, assumed to be present in the calculations above, is violated, in 
addition to TR symmetry. 
The presence of at least one of those symmetries, i.e. TR or inversion, is necessary for the existence of the BCS superconducting state, since 
only those symmetries guarantee that the band eigenstates at momenta $\bk$ and $- \bk$ have the same energy. 
In a Weyl metal, either TR or inversion must be violated, to remove the two-fold Kramers degeneracy. 
In the model that we have discussed above, TR was violated from the start. This already puts BCS-type superconductivity under some strain, 
which is manifest in the gap function having nodes. 
When inversion symmetry is violated as well, the Weyl nodes will generally be shifted to different energies, which implies that the states with momenta 
$\bk$ and $-\bk$ will have different energies. Once the inversion breaking is strong enough, such that the energy difference between the Weyl nodes is comparable 
to the BCS pairing amplitude $\Delta$, given by Eq.~\eqref{eq:36}, the BCS state will be destroyed (possibly going through an intermediate ``ordinary" small-wavevector 
FFLO state, which we will not discuss here).  
However, the FFLO state is largely unaffected by this, since its existence relies not on the exact microscopic inversion symmetry, but rather on the low-energy 
chiral symmetry, which is unaffected by the broken inversion symmetry, unless inversion breaking is so strong that the energy difference between the Weyl nodes 
becomes comparable to the Fermi energy.
Thus, provided the inequality Eq.~\eqref{eq:56} may be satisfied (which, as discussed above, appears to be hard in the weak coupling regime, but may be possible at intermediate coupling), the FFLO state will be realized. 
However, it is important to keep in mind that since nonlinear chiral symmetry violating corrections to the band dispersion are always present in a real Weyl metal,
realizing the FFLO state certainly requires a finite pairing interaction strength, as the logarithmic divergence of the FFLO pairing susceptibility will always be cut off by the 
nonlinearity. 
From this viewpoint, a Weyl metal, in which both TR and inversion are violated, is an example of a metal without any weak-coupling BCS instability and thus 
with a true Fermi liquid ground state at $T = 0$. 
Another closely related and interesting possibility, worthy of further, more detailed study, is a state, in which the superconducting order parameter is nonzero on one of the Fermi surface sheets, but is zero on the other one, since the inequality Eq.~\eqref{eq:56} may be satisfied for one of the sheets, but not the other, when the inversion symmetry is absent. 
This would be a {\em helical superconductor},~\cite{Samokhin94,Samokhin14} coexisting with a normal Fermi liquid.
Helical superconductors have been studied extensively in the general context of superconductivity in noncentrosymmetric materials,~\cite{Samokhin08}
and their realization in Weyl metals would be of significant interest. 

Our results may be used as a starting point for further studies of superconductivity in Weyl metals.
In particular, since the nodal BCS state is topologically nontrivial, as discussed in Section~\ref{sec:4} and in Refs.~\onlinecite{Meng12,Tanaka15}, the question of its electromagnetic response seems to be of interest. 
\begin{acknowledgments}
We acknowledge financial support from Natural Sciences and Engineering Research Council (NSERC) of Canada (GB and AAB) and from Swiss NF and NCCR QSIT (AAZ). 
\end{acknowledgments}
\bibliography{references}

\end{document}